\documentclass[article]{aa}

\usepackage{natbib}
\bibpunct{(}{)}{;}{a}{}{,} 
\usepackage{graphicx}
\usepackage{txfonts}

 \usepackage{hyperref}
\usepackage{multirow}

\newcommand{\euclid}{\textit{Euclid}\xspace}

\usepackage{algorithmicx}
\usepackage{algorithm}
\usepackage{algpseudocode}
\begin{document}

   \title{Detecting unresolved binary stars in \euclid VIS images}

   \subtitle{}

   \author{T. Kuntzer          
          \and
          F. Courbin
          }

   \institute{Institute of Physics, Laboratory of astrophysics, Ecole Polytechnique F\'ed\'erale de Lausanne (EPFL), Observatoire de Sauverny, CH-1290 Versoix, Switzerland 
   \\\email{thibault.kuntzer@epfl.ch}}

   \date{Received 16 March 2017; accepted 22 August 2017}

\abstract{Measuring a weak gravitational lensing signal to the level required by the next generation of space-based surveys demands exquisite reconstruction of the point-spread function (PSF). However, unresolved binary stars can significantly distort the PSF shape.
In an effort to mitigate this bias, we aim at detecting unresolved binaries in realistic \euclid stellar populations. We tested methods in numerical experiments where (i) the PSF shape is known to \euclid requirements across the field of view, and (ii) the PSF shape is unknown.
We drew simulated catalogues of PSF shapes for this proof-of-concept paper. Following the \euclid survey plan, the objects were observed four times. We propose three methods to detect unresolved binary stars. The detection is based on the systematic and correlated biases between exposures of the same object. One method is a simple correlation analysis, while the two others use supervised machine-learning algorithms (random forest and artificial neural network).
In both experiments, we demonstrate the ability of our methods to detect unresolved binary stars in simulated catalogues. The performance depends on the level of prior knowledge of the PSF shape and the shape measurement errors. Good detection performances are observed in both experiments. Full complexity, in terms of the images and the survey design, is not included, but key aspects of a more mature pipeline are discussed. Finding unresolved binaries in objects used for PSF reconstruction increases the quality of the PSF determination at arbitrary positions. We show, using different approaches, that we are able to detect at least binary stars that are most damaging for the PSF reconstruction process.}

   \keywords{Methods: data analysis -- Methods: statistical -- (Stars:) binaries (including multiple): close}

   \maketitle

%
   \section{Introduction}
   \defcitealias{Kuntzer2016a}{KC16}
   
Future space-based observatories such as \euclid\footnote{\url{http://www.euclid-ec.org/}} \citep{Euclid} and {\it WFIRST} \citep{WFIRST} require exquisite point-spread function (PSF) stability and measurement to achieve their scientific goals, in particular in view of the weak-lensing cosmological probe \citep{Cropper2013, Massey2013,Schneider2006}. Several instrumental effects have been identified as nuisances in the PSF determination process, such as the colour of the object and colour gradients across the objects \citep[][]{Voigt2012, Semboloni2013}, the brighter-fatter effect \citep{Antilogus2014}, or detector distortions that are due to temperature gradients.

Other astrophysical or atmospheric nuisances also contribute to the degradation of the signal, such as satellite trails, cosmic rays, stray light \citep[e.g.][]{Desai2016}, or PSF distortions by unresolved binary stars \citep[][KC16 hereafter]{Kuntzer2016a}. 
We showed in \citetalias{Kuntzer2016a} that unresolved binaries can significantly alter the shape of the \euclid PSF, even if many (thousands of) stellar images are used together to reduce this effect. 

In this paper, we propose a technique to identify unresolved binary stars that may hamper a proper PSF measurement. We base our method on the systematic nature of the deformation induced by binary stars. While the deformation of individual stellar images is of the order of the instrumental noise, correlations between different exposures of the same object can lead to the detection of a binary star, or at least to flagging an object as non-reliable for subsequent PSF determination.

A similar approach was used by \citet{Hoekstra2005} to detect eclipsing binaries in the Optical Gravitational Lensing Experiment (OGLE).
Detecting unresolved binaries from the ground using this approach has already proven successful \citep{Terziev2013,Deacon2017}.
We build on the technique pioneered by \citet{Hoekstra2005} but apply more sophisticated algorithms, including artificial neural networks (ANN). We also study the applicability of our techniques to the specific case of \euclid, with the goal of identifying unknown and unresolved binaries that affect the quality of the PSF reconstruction, as described in \citetalias{Kuntzer2016a}.

The observing strategy of \euclid is to take four dithered exposures \citep{Cropper2016} of each field for the weak-lensing surveys. 
The deep fields, located near the galactic poles, will be repeatedly observed over the course of the mission, at different orientations. 
These mutiple images of the same stars can be used, when combined, to reconstruct an estimation of the PSF to account for its effect. 
They can also be individually used to check for systematic biases in the shape estimates of the light profiles. Flagged objects can be removed from the stellar catalogues used in view of the PSF reconstruction.

In this proof-of-concept paper, we choose to work at the catalogue level, and not to measure \euclid-simulated images of single or multiple images.
Our choice to work at the catalogue level is deliberate. Our goal is to show that our approach can be an efficient method to flag unresolved binaries when a catalogue of stars with shape measurements is available. 
We do not aim at building a complete PSF shape measurement and reconstruction pipeline.

We show on simulated catalogues of non-dithered observations that we can robustly detect unresolved binary stars in a sample of point sources. This article is organised as follows: we detail the algorithms and associated performance metrics in Sect.~\ref{sec:algos}. 
Section~\ref{sec:binaries} presents a summary of the statistical knowledge of the binary systems, and Sect.~\ref{sec:mock_obs} discusses the mock observations of stars with the VIS instrument.
We then illustrate the performance of the algorithm with image simulations in Sect.~\ref{sec:results}. Finally, Sect.~\ref{sec:summary} summarises our findings.

%
%
\section{Definitions, scheme, and training data}  
\label{sec:algos}

 We propose different algorithms that learn the difference between systematic and stochastic distortions to predict whether an object is a single or a binary star. The examples used for training (in the machine-learning sense) could be drawn from real data of sources whose binary nature is known or from simulated data.

\subsection{Effect of binaries on the PSF shape parameters}
A more in-depth discussion of the effect of binaries on the shape of the PSF is provided in \citetalias{Kuntzer2016a}.
In this paragraph, we summarise the main results of \citetalias{Kuntzer2016a}
very briefly.
Throughout, we use the term contrast for the difference in magnitude $\Delta m$ between the host (or main) and companion star $\Delta m = m_\text{companion} - m_\text{host}$.
The presence of a companion can change the measured PSF ellipticity by an order of 1\% when the angular separation between host and companion is $r=0.\mkern-4mu^{\prime\prime}01$ (i.e. a tenth of the \euclid pixel).
Similarly large effects are seen on the size measurement.
The distortion of the PSF induced by a companion star depends on the angular separation and on the contrast in magnitude \citetalias{Kuntzer2016a}. 
The expected distortions of the PSF caused by binary objects are systematic in nature.
On average, the deformations due to binary stars tend to make the PSF rounder and larger.
On a case-by-case basis, however, the deformation of the PSF may not make it rounder, but could increase the ellipticity.

The deformation of the PSF measured on the image of a binary system depends on the relative positions of the host and the companion, and on the magnitude contrast.
For the wide-field survey of the \euclid mission, the exposures are taken in a sequence, thus the physical properties  describing a binary system will not change.
In the deep exposures, taken in a much longer time interval, the properties of the binary system may change.

\subsection{Definitions}

For simplicity, we assume that a given star is imaged in several exposures at similar spatial positions on the detector. The observables used to classify an object are the shape parameters of the PSF. These observables, which are impervious to the measurement method, are the complex ellipticities ($e_1$ and $e_2$) and the PSF ``size'' $R^2$.
These can be defined using quadratic moments as follows.

Let $I(\pmb{\theta})$ be the brightness distribution of an an image. $\bar{\pmb{\theta}}$ would be the position of the centroid.
The tensor of second-order moments $q_{jk}$ computed along the first and second axis of an image is given by
\begin{equation}
  q_{jk}=\frac{\int \left(\theta_j-\bar{\theta}_j\right)\left(\theta_k-\bar{\theta}_k\right) \mathrm{I}(\pmb{\theta}) \mathrm{d}^2\theta}{\int \mathrm{I}(\pmb{\theta}) \mathrm{d}^2\theta},\quad j,k\in\{1, 2\}.
\end{equation}
In a noise-free image, the complex ellipticity components of a light profile are obtained from \citet{Kaiser1995}:
\begin{equation}
  [e_1, e_2] =\left[\frac{q_{11} - q_{22}}{q_{11} + q_{22}}, \frac{2q_{12}}{q_{11} + q_{22}}\right].
\end{equation}
The size of the PSF is defined as
\begin{equation}
  R^2 = q_{11} + q_{22}.
\end{equation}
The standard deviations of the variations in the PSF for the \euclid survey must be smaller than $\sigma(e_i) \leq 2\times 10^{-4}$ for the ellipticity components and $\sigma(R^2)/\langle R^2\rangle \leq 1\times10^{-3}$ for the size according to science requirement 4.2.1.4 \citep{EuclidSciReq,Paulin-Henriksson2008,Euclid}. Each \euclid image will contain from 2\,000 to 3\,000 stars \citep{Cropper2013,Euclid}, depending on Galactic latitude. If the PSF reconstruction scheme is optimal, the knowledge on each individual star is $\sqrt{2\,000}\approx50$ worse than the requirements, that is, 1\% for the ellipticities and 5\% for the size on each individual star. 

\subsection{Simulated data}
\label{sec:simdata}
We stress that we work at the catalogue level and not at the pixel level. We assume that the PSF can be measured to within the \euclid expected performances (as stated in the previous paragraph) for each individual star.
From a catalogue, we can flag suspicious observations of what has been considered as a single star.
The precision and accuracy we take on the shape parameters are given by the \euclid requirements. 
The preparation of a PSF pipeline that meets the \euclid requirement is beyond the scope of this paper.
For this reason, we chose to work with a catalogue that fulfils the claimed requirements. 
In this way, our results do not depend on any given shape measurement method either. 

For the observations of single stars, we assume that the observed PSF parameters are noisy around their fiducial value (we discuss this assumption in Sect.~\ref{sec:assumptions}). 
In the case of a binary system, however, the observables $e_1, e_2$, and $R^2$ are systematically modified by a value that depends on the angular separation and the contrast of the system.
We assume throughout that the PSF parameters $e_1,e_2,\text{a}\text{nd
}R^2$ are measured to the above precision, that is, 1\% (elipticity)
and 5\% (size) per star observation.

Catalogues of simulated single and binary system observations are produced by the following procedure and described in more details in Sect.~\ref{sec:mock_obs}:

\begin{enumerate}

\item Draw a stellar population from the Besan\c{c}on Galaxy Model \citep[BGM,][]{Robin2003} for a \euclid-like sky.

\item Using empirical knowledge on the distribution of binary stars \citep{Duchene2013}, compute the confidence that any given star is a binary system and compute its characteristics (semi-major axis of the orbit and mass ratio).

\item Draw a realisation of the binary population with selections in angular separation and contrast.

\item Estimate the observed PSF parameters for each object in the catalogue and for each exposure. 

\end{enumerate} 

\subsection{Principle of the binary detection algorithms}
\label{sec:principle}
\begin{figure}[!h]
 \begin{center}
  \includegraphics[width=0.9\linewidth]{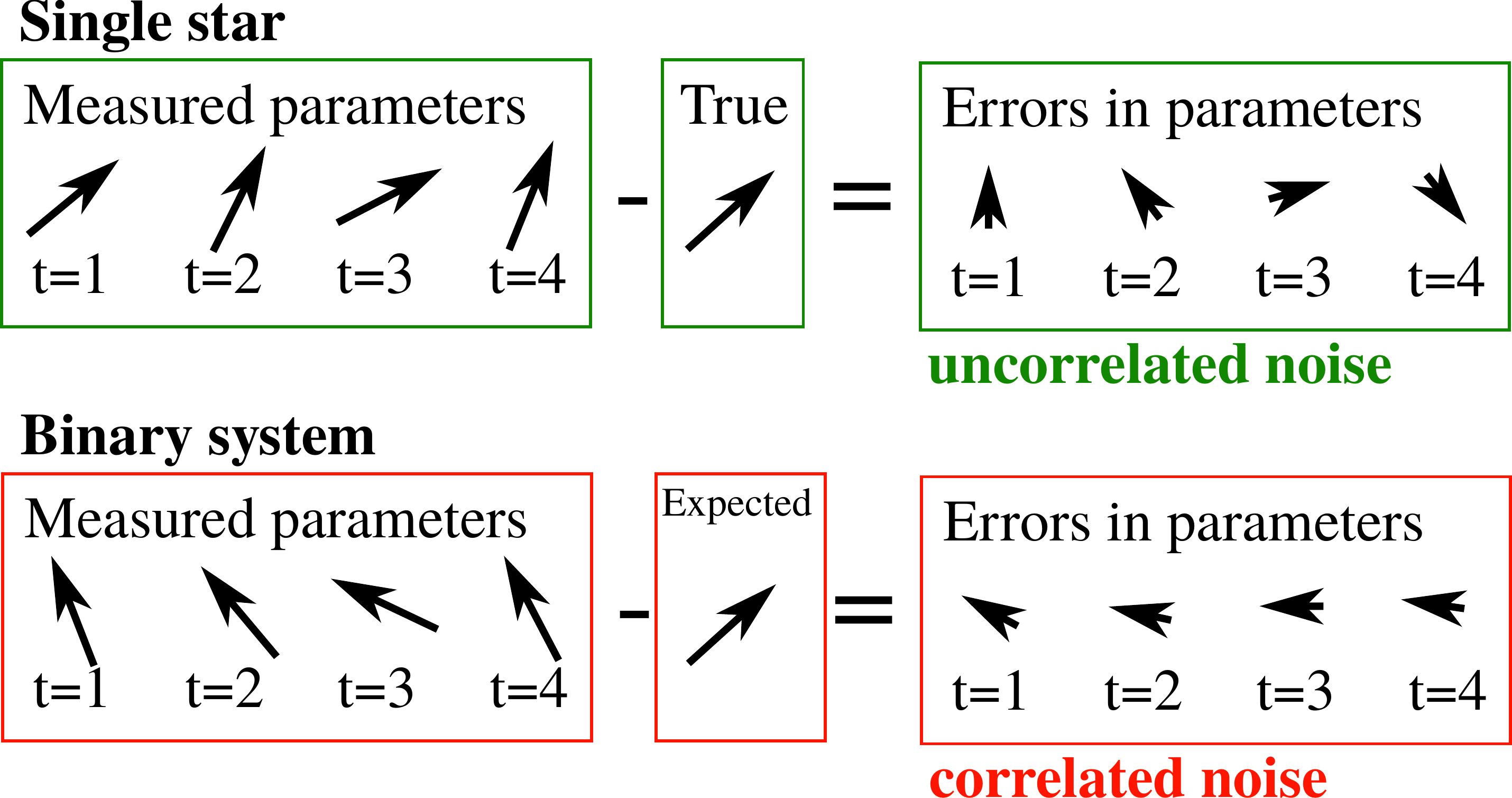}
 \end{center}
 \caption{\label{fig:correlation} Sketch of the measurement principles. The four observations are compared to the fiducial value of the shape parameter. A binary star will affect the shape of the PSF systematically across the four exposures (bottom), whereas the measurement errors are uncorrelated (top). }
\end{figure}
Our detection algorithms of unresolved binaries follows this procedure and is illustrated in Fig.~\ref{fig:correlation}.
\begin{enumerate}
\item Compute the relative error, $\delta^{(i)}$, based on the PSF parameters $p^{(i)} =e_1^{(i)}, e_2^{(i)}$ or $(R^{2}){^{(i)}}$ for a star $i$ with respect to a fiducial value $p^{(i)}_0$. The relative error is
\begin{equation}\label{eq:relerr}
\delta^{(i)} = \frac{p^{(i)}-p^{(i)}_0}{p^{(i)}_0}. 
\end{equation}
For each star $i$, four such relative errors are measured for
each parameter because there are four observations. 
We denote access to the relative error in  exposure $t$ as $\delta^{(i)}(t)$.
The fiducial parameter fields are either known {\it \textup{a priori}} or are directly inferred from the measurement PSF parameters using an iterative field estimation and outlier removal algorithm. 

When the PSF field is unknown, the PSF at a given point is reconstructed using the parameters of neighbouring single stars. The interpolated value can be interpreted as the fiducial value at that point. When the measured relative errors are too different from the fiducial value, the object is considered as an outlier and is
removed from the list of single stars.
This process of interpolating a fiducial value based on neighbouring single stars and marking it single or outlier is repeated several times. This algorithm is detailed in Sect.~\ref{sec:uknPSF}.
\item Use the resulting relative errors $\delta^{(i)}$ as the features for the algorithms to infer the presence of binaries.

\end{enumerate} 

\section{Algorithms to identify unresolved binary stars}
\label{sec:binfind}

We propose three different classifiers that yield a binary output (binary star or single star) or, similarly, a number between 0 and 1 that translates the confidence of detection. The three classifiers tested here are (i) a simple auto-correlation of the input features, and two supervised-learning methods, namely (ii) random forests (RF) and (iii) ANNs. All three methods are evaluated on two binary system selection criteria: their (i) angular separation, and (ii) contrast.

%
\subsection{Algorithm 1: Auto-correlation function} 
\label{sec:method_acf}
 
 \begin{figure}[!h]
 \begin{center}
  \includegraphics[width=0.9\linewidth]{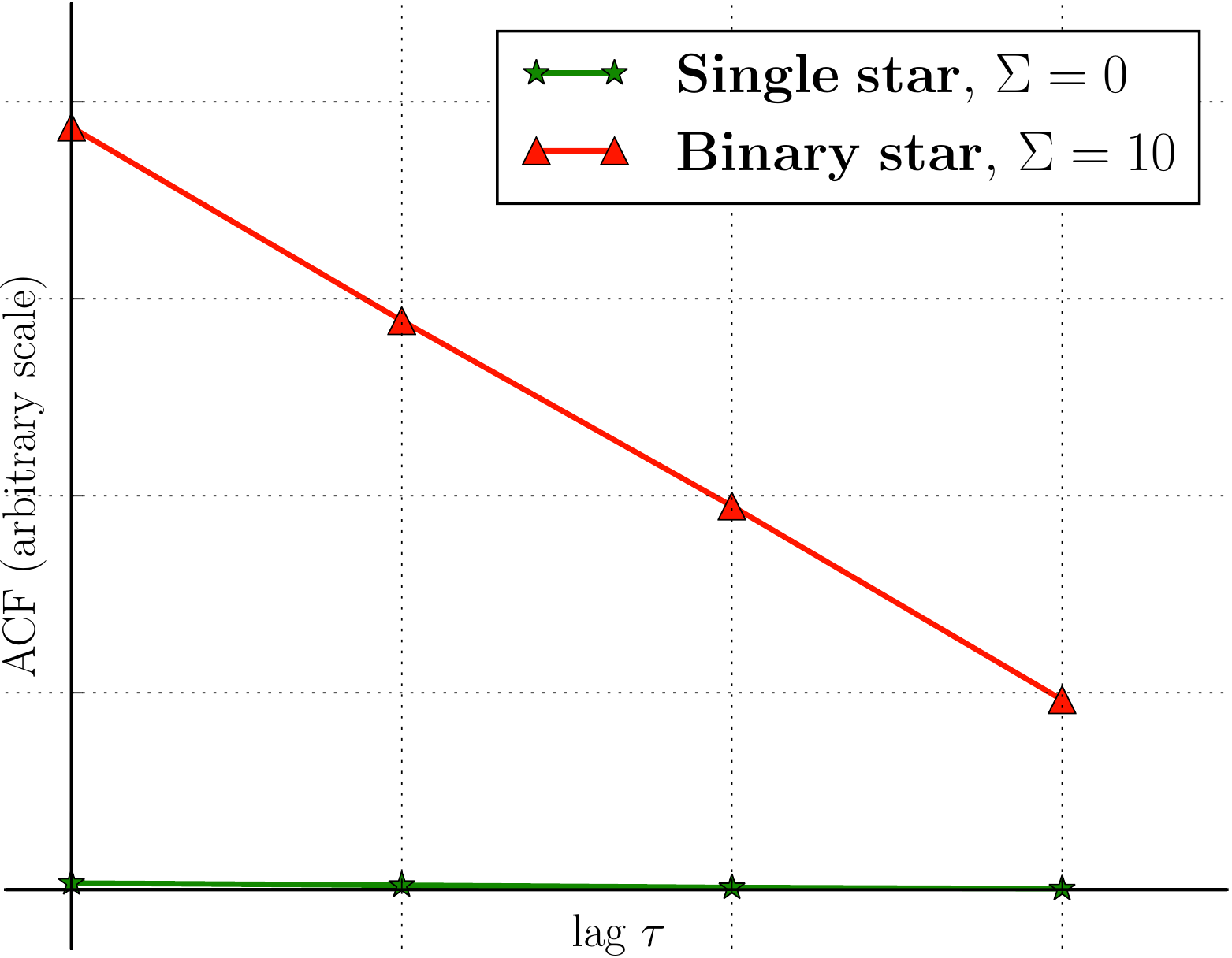}
 \end{center}
 \caption{\label{fig:acf_sketch} Illustration of the result of the ACF for a relative error $\pmb{\delta}$ of a given shape parameter.
 We show the ACF for a single star (green) and a binary system (red).
 For simplicity, the noise level in the shape measurement is set to unrealistically low values.
 The shape parameters of the single stars are uncorrelated between exposures, thus the ACF coefficients are all zero, yielding the scalar quantity $\Sigma=\Sigma_{\tau=0}^3 \mathrm{ACF}(\tau)=0$. The shape parameters are correlated for a binary system, and a strong auto-correlation between exposures is observed, with a high scalar quantity $\Sigma=10$. }
\end{figure}

For each object $i$ in the field, we consider a list of four relative errors $\delta^{(i)}$ for each of the shape parameters. These lists can be understood as time series of the relative errors. 
As described in Sect.~\ref{sec:principle} and illustrated in Fig.~\ref{fig:correlation}, when the object is a single star, the errors made on the measured parameters are stochastic.
However, when the object is a binary system, the errors are correlated because the relative position of the main and companion star do not change between the first and last exposure.
A measure of the correlation of the signal with itself can distinguish between no correlations (low auto-correlation score) or correlations (high auto-correlation score), which acts as a proxy to the binary or single star nature of the object.
 
An estimate of the auto-correlation function (ACF) coefficient for a lag of $\tau$ for a vector of relative errors $\pmb{\delta}$ observed at $t=\{1,2,3,4\}$ is given by
\begin{equation} \label{eq:acf}
 \mathrm{ACF}(\tau) =\frac{1}{ (n-1)\sigma^2}\sum_{t=1}^{n=4} [\delta{(t)} - \bar{\delta}][\delta{(t+\tau)} - \bar{\delta}], \,
\end{equation}
where $\sigma^2$ is the sample variance, and $\bar{\delta}$ is
its mean value as computed from $n$ observations of $\delta$. $n=4$ exposures in our work. For simplicity, we dropped the star index $i$ in the above equation. This method is data driven: it is based on the assumption that the relative errors are more strongly correlated for binary systems than for single systems. No physics is included in the model.
We sketch the result of the ACF method for a binary system and a single star in Fig.~\ref{fig:acf_sketch}.

We sum the four estimated ACF coefficients for the four observations to create a scalar quantity $\Sigma=\sum_{\tau=0}^3\mathrm{ACF}(\tau)$.
We choose a value for a separating threshold that classifies highly auto-correlated observables as binary stars and, conversely, links low auto-correlation of the observables to single stars.
Each PSF parameter yields an estimate of the auto-correlation. We refer to each of these classifications as $e_1,e_2$, and $R^2$ channels.
The outputs of the three detection channels can be combined to improve the classifiers' performance. 
A three-channel classifier is obtained by taking a weighted average of the output of each individual channel. 
The weights are empirically determined by evaluating the individual channel's performance.
The prediction of the three-channel classifier is again compared to a separating threshold. 
The ACF classifier is run against a training dataset to determine the separating threshold, which is set such that the false-positive rate is 10\%.

\begin{figure*}
 \begin{center}
  \includegraphics[width=0.45\linewidth]{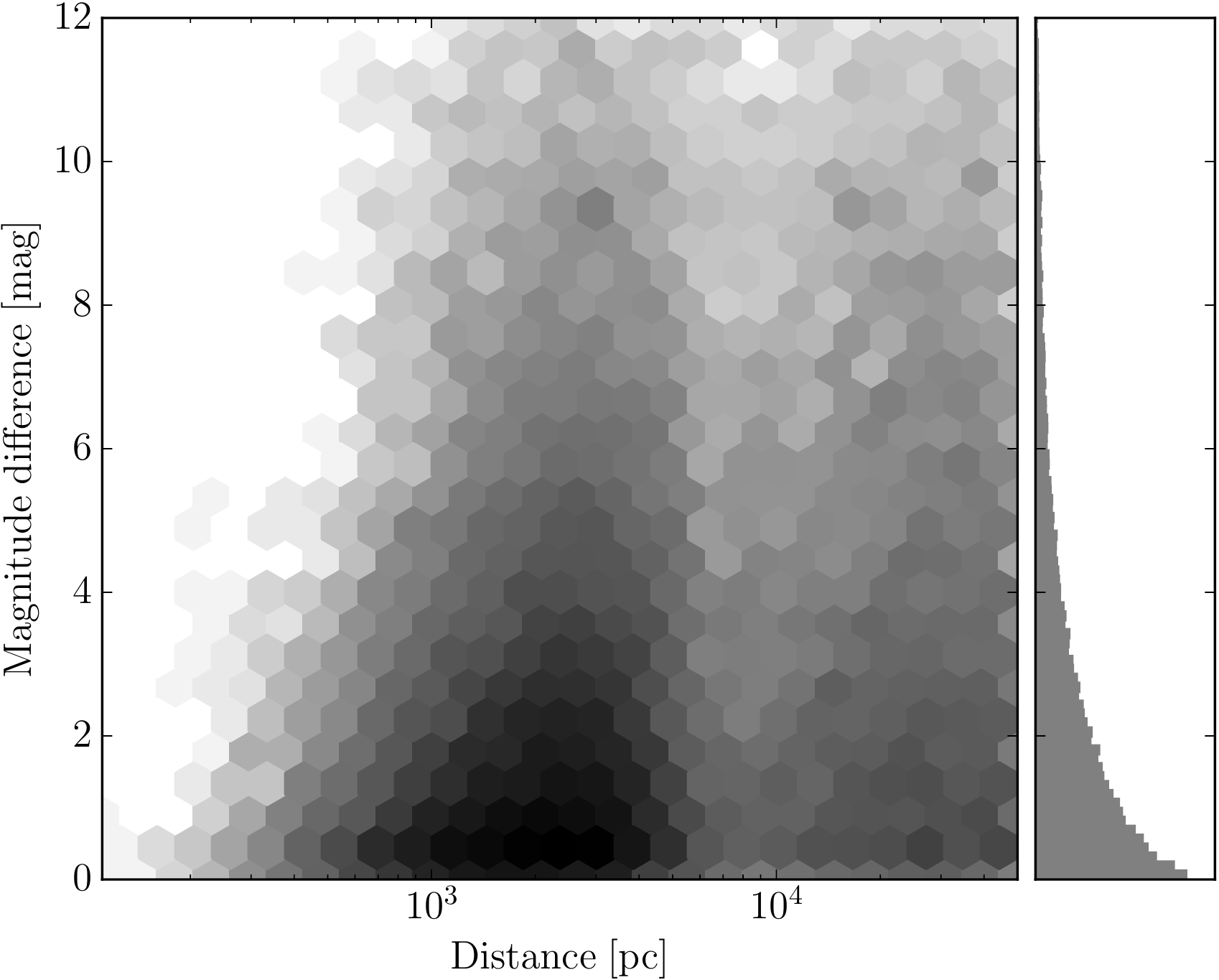}
  \includegraphics[width=0.45\linewidth]{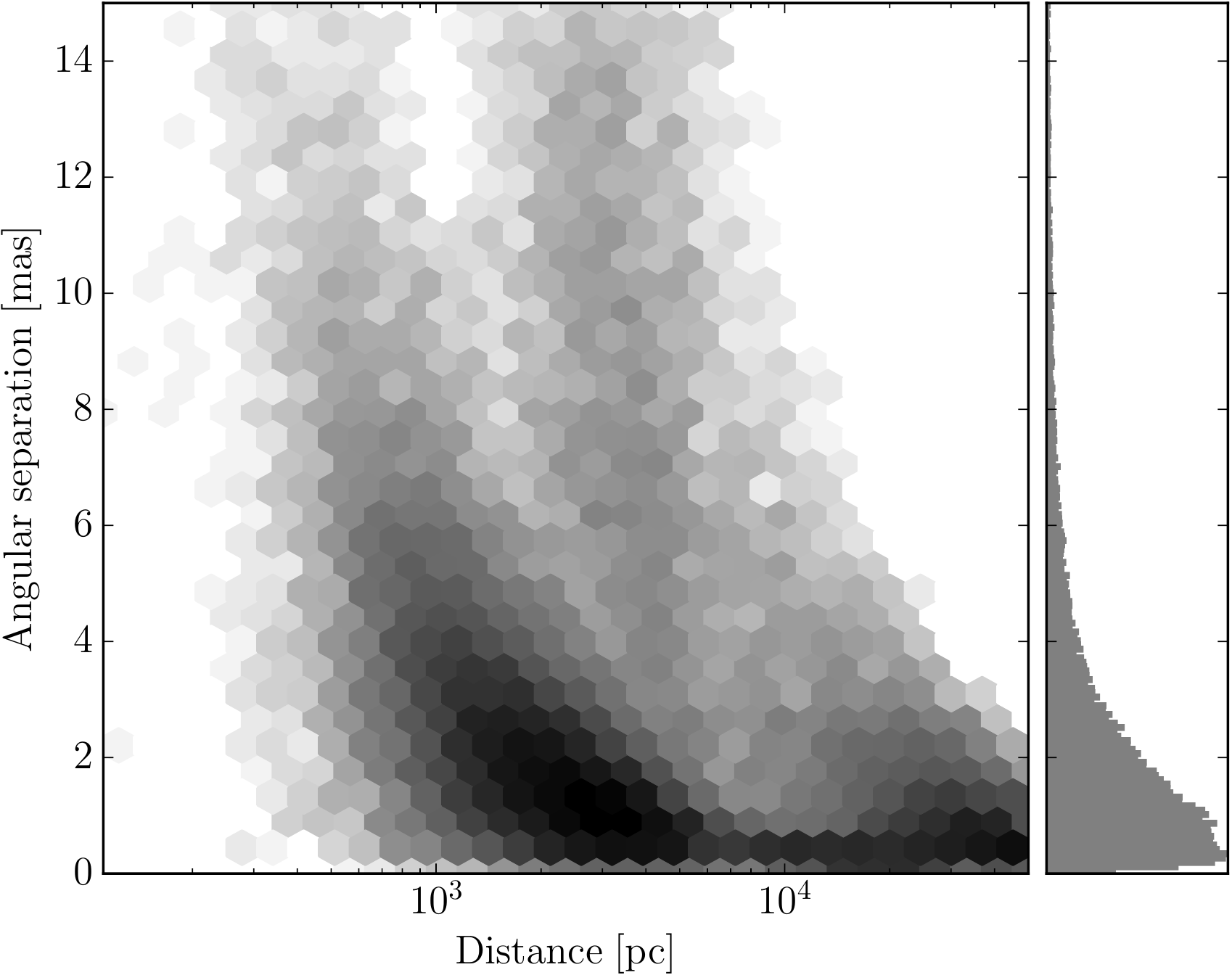}
 \end{center}
 \caption{\label{fig:stellar_pop} Numerical model of the binary star density as a function of distance to Earth and contrast (left) and angular separation (right). The grey scales and the histograms are linear and relative to the maximum bin. We select stars in the magnitude range $18\leq I(AB)\leq24.5$. Both plots exhibit bimodal distributions with low-mass main stars located at shorter distances. More massive stars are not located in our direct neighbourhood, as suggested by the distribution of angular separation with distance. Companions of very low mass stars (down to the brown dwarf limit) are rare with the applied selection cuts. }
\end{figure*}

%
\subsection{Algorithm 2: Random forest}

When the expected distortions of the shape parameters are known, we can train an algorithm to recognise patterns in the data.
The random forest (RF) algorithm is a large ensemble of simple but weak decision trees. 
Decision trees are a class of non-parametric supervised-learning classifier or regression methods that hierarchically approach the data. 
The training of a decision tree consists of learning a set of decision rules based on the features and the ground truth \citep{Breiman2001}. 
Decision trees taken individually tend to overfit the data. Averaging over a large number of trees -- a forest -- reduces the classification error.

In the same way as for the ACF classifier, the features adopted for the RF are the relative errors made on the PSF parameters as given by Eq.~(\ref{eq:relerr}). The ground truth is the nature of the object (binary or single star). 
The depth of the individual trees is not limited, and a forest of 50 trees is constructed.

We used the RF implementation of the {\tt Python} package {\tt scikit-learn} with the Gini impurity criterion \citep{scikit-learn}. 

%
\subsection{Algorithm 3: Artificial neural networks}

Feed-forward artificial neural networks (ANNs) belong to a class of supervised algorithms that can be used for regression or classification tasks. We applied ANNs to classify binary and single stars. For an introduction to ANNs, we refer to \citet{Bishop1995}. The ANNs are made of an input layer, an arbitrary number of hidden layers, and an output layer. The hidden and output layers consist of a number of neurons that take as input a vector $\pmb{x}$ and return a scalar $h(\pmb{x},\pmb{w}, b)$ through
\begin{equation} \label{eq:ANN}
 h(\pmb{x},\pmb{w}, b) = h\left(\sum_{i=1}^N w_i x_i + b\right),
\end{equation}
where $\pmb{w}$ and $b$ are the weights and the bias, respectively. The monotonic and continuous function $h$ is the so-called activation function. 
In the following, we use the hyperbolic tangent as activation function. 
The ability of an ANN to find patterns shared by the the input data (the features) and the labels of each image (the ground truth and its prediction) depends on the architecture of the network, in that case, the number of neurons and layers. 
The architecture must be optimised for specific applications.
We found that for our application, three layers of 15 neurons each performs well.
We trained the network on a standard mean-square-error cost function.
The size of the training and test datasets is given in Sect.~\ref{sec:res_ann}.

The choice of the ANN implementation, like the choice of the RF implementation, is arbitrary from the outset. A comparison
of the details of each implementation is beyond the scope of this paper. We used the {\tt Python} bindings for the Fast Artificial Neural Network Library ({\tt FANN}) implementation and its optimiser described in \citet{Nissen2003}.

The ACF and the RF methods were applied to the simulated data described in Sect.~\ref{sec:results}
The performance of the ANNs is analysed in Section~\ref{sec:res_ann} on worst-case experiments only, namely on an unknown field in
both stellar population and PSF field parameters. 

%
\subsection{Metrics for the detection performance}
\label{metrics}

We used three metrics to quantify the effectiveness of detecting binary stars:

\begin{itemize}
\item The \emph{\textup{receiver operating characteristic}} (ROC) is a graphical representation of the performance when the decision threshold is varied. The abscissa represents the false-positive rate (FPR, or fall-out),
\begin{equation}
 \mathrm{FPR} = \frac{\mathrm{false\ positives}}{\mathrm{false\ positives + true\ negatives}},
\end{equation}
while the ordinate encodes the true-positive rate (TPR, also called sensitivity and recall),
\begin{equation}
 \mathrm{TPR} = \frac{\mathrm{true\ positives}}{\mathrm{true\ positives + false\ negatives}}.
\end{equation}
An algorithm that would randomly classify a star as a binary lives on the diagonal FPR$=$TPR. 
An ideal classifier would be represented by a curve passing through the coordinate $[$FPR$=$0, TPR$=$1$]$. The ROC allows for a quick and reliable comparison between different classifiers, marginalising over the detection threshold. The choice of the separating threshold can be based on the ROC by trading off the TPR for the FPR \citep{Kleinbaum2010}.
\item The \emph{area under the curve} (AUC) is the integral of the ROC over the FPR. It summarises the ROC as a single scalar. A random algorithm would score $\text{AUC}=0.5,$ while an ideal method would reach $\text{AUC}=1$. For this metric to be high, the true positive rate and the precision (the ratio of the true positives to the sum of true and false positives) must both
be high.
\item The $F_1$\emph{ score} is a scalar metric widely used for binary classification, defined as
\begin{equation}
 F_1 = \frac{2\mathrm{TP}}{2\mathrm{TP}+\mathrm{FN}+\mathrm{FP}},
\end{equation}
 where $\mathrm{TP}$, $\mathrm{FN}$, and $\mathrm{FP}$ are the numbers of true-positive, false-negative, and false-positive classifications, respectively \citep[e.g.][]{Herlocker2004}. 
This metric is not only sensitive to the number of correctly detected binaries (TP), but also to the number of binaries classified as single stars (FN) and inversely (FP). A perfect classification would yield $F_1 = 1$, while a completely random decision-making algorithm would have an $F_1$ score that tends to zero. 
\end{itemize}
%

%
\section{Binary star population in \euclid} 
\label{sec:binaries}

\begin{figure}
 \begin{center}
  \includegraphics[width=1.\linewidth]{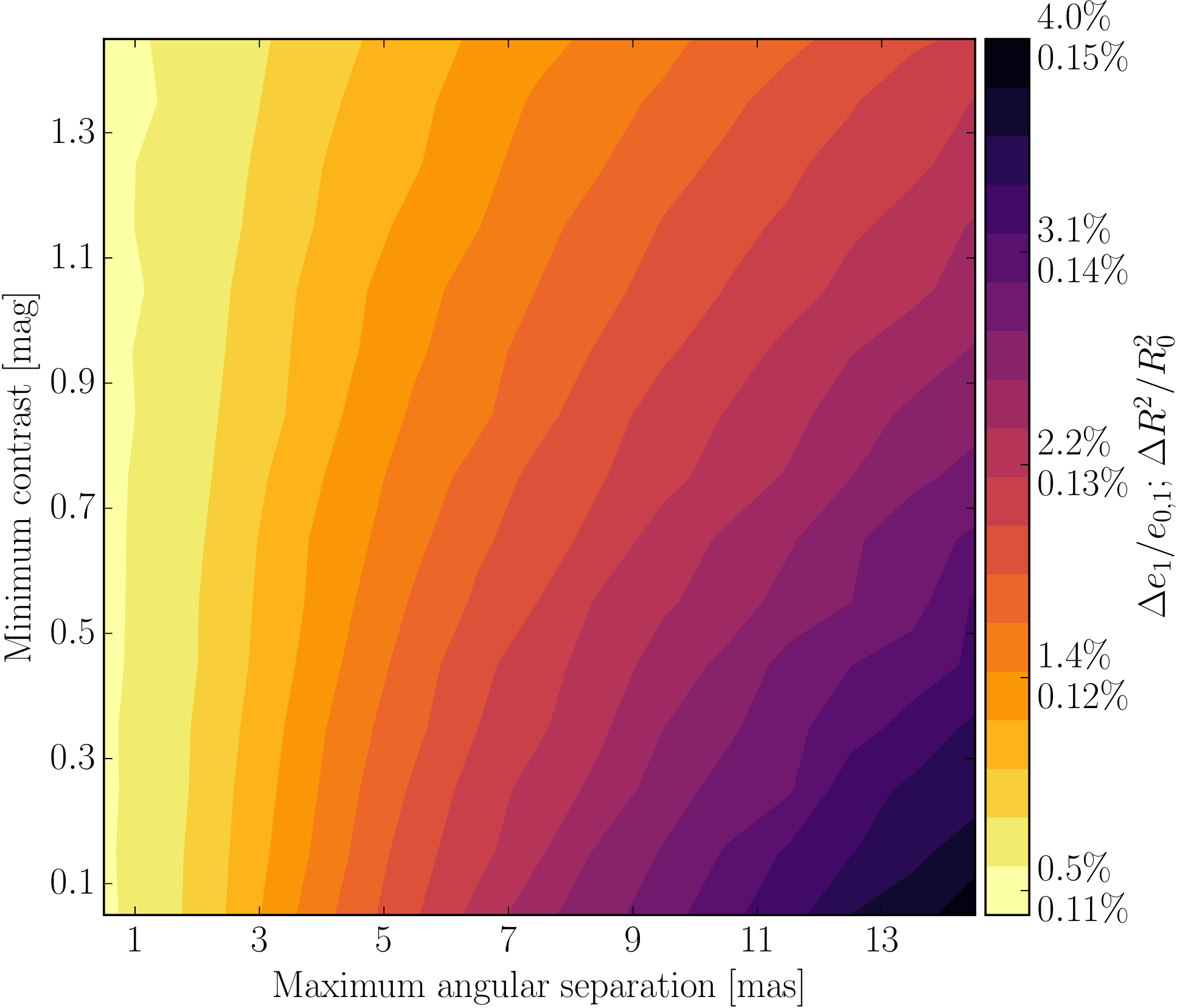}
 \end{center}
\caption{\label{fig:dang_star} 95-percentile relative errors of the PSF shape parameters $e_1$ and the size. The relative errors in $e_2$ are about twice smaller. The values of the relative errors are given in the colour bar for the ellipticity (top number) and for the size (bottom number). The magnitude selection is $18 \leq I(AB) \leq 24.5$. The distributions of relative errors contain all binaries that have at most an angular separation on the $x$-axis and at minimum a contrast on the $y$-axis.}
\end{figure}

In order to test our binary star finders, we designed \euclid-like simulated images, building on a stellar population with realistic binary fractions. The latter as well as the orbital parameter distributions and luminosity contrasts have been studied in volume-limited searches. 
Early searches were restricted to low numbers of stars, hence leading to incomplete samples \citep[e.g.][]{Abt1976,Marcy1989,Duquennoy1991}. More recent searches harvested more data, but the total sample of stars with a detailed characterisation is still limited to a few hundred
\citep{Raghavan2010,Milone2012,Cummings2014,Nardiello2015,Riddle2015,Rodriguez2015}.

\citet{Duchene2013} compiled the available data and proposed distributions of multiple systems in terms of multiple star fraction, mass ratio to the main star, and angular separation. 
We used these parameters to a create stellar population based on BGM realisations and with realistic binary fractions.
Figure~\ref{fig:stellar_pop} shows the contrast and the angular separation of binary stars as a function of distance to Earth for stars located in the direction of the anti-Galactic centre, with Galactic latitude in the range $+15^\circ<b<+75^\circ$. 
We selected stars when their apparent magnitude was in the range of the expected magnitude for the \euclid wide-field survey (i.e. $18 \lesssim I(AB) \lesssim 24.5$). 
The bimodal nature of the distributions in Fig.~\ref{fig:stellar_pop} is explained by the mass of the main star. Massive stars in the
selected magnitude range are rare in the neighbourhood of the Sun, and inversely, low-mass stars are not selected by our cut in magnitude at large distances.
Figure~\ref{fig:stellar_pop} also presents histograms of the stars as a function of contrast and angular separation.
They demonstrate that most binary systems that will be seen by \euclid have a low contrast, meaning that both stars are similar in terms of magnitude and have a very small angular separation, with about 75\% of the binaries separated by less than $3.5$~milliarcseconds (mas) and 50\% below $1.7$~mas. 
Most companions have a magnitude similar to the main star of the system, but a wide diversity of contrast is expected. 
Systems that contain brown dwarf companions are also predicted.

Figure~\ref{fig:dang_star} illustrates the magnitude of the relative errors due to binaries at small separations and low contrast. 
We show the 95-percentile value of the expected relative errors in the shape of PSFs over a given binary population.
The intuition that larger relative errors occur at larger angular separation and low contrast is confirmed. 
Quantitatively, the relative errors follow the same trends, but the effect is strongest on the $e_1$ component, followed by $e_2$, whose relative errors are about half as important. 
The relative errors induced on the size are an order of magnitude weaker than on $e_2$. Since the size is also predicted to be measured five times less accurately than the ellipticities (see Sect.~\ref{sec:simdata}), the latter are more sensitive indicators of the presence of binaries than the former. 

The change in PSF shape due to the presence of binary stars is not yet budgeted in \euclid, and there is no requirement available so far on the performance of any binary rejection method. 
However, Fig.~\ref{fig:dang_star} indicates a detection limit of $\sim3$~mas in separation almost independently of contrast. 
This ensures that the distortions due to binaries are of the same order of magnitude as the measurement error per star for the ellipticity, that is, $\sim1\%$.

%
\section{Mock catalogues of PSF shape parameters} 
\label{sec:mock_obs}

In this section, we explain how we create the catalogues of PSF parameters for any given single and binary system at any spatial position in the four central CCDs of the \euclid VIS detector.

For the shape parameters contained in the catalogues to be in a realistic range, we measured super-sampled noise-free \euclid-simulated PSFs.
We simulated 600 \euclid PSFs at randomly selected positions on the detector using the ray-tracing tool {\tt Zemax} \citep{NgoleMboula2015}. The images of these PSFs are super-sampled by a factor of 12 (i.e. the pixel is one-twelfth of $0.\mkern-4mu^{\prime\prime}1$).
The shapes of the simulated single and binary stars were measured using the adaptive-moment scheme proposed by \citet{Bernstein2002} and based on re-Gaussianisation \citep{Hirata2003} as implemented in {\tt Galsim} \citep{Rowe2015}.
To obtain a realistic stellar population, we used the BGM, with the \euclid cuts on magnitude ($18 \lesssim I(AB) \lesssim 24.5$) combined with the information on the estimates of the fraction of multiple stars and orbital parameter distributions derived from \citet{Duchene2013}. 
We restricted the stellar population to astrophysical binaries. Coincidental binaries of field stars are negligible given the star density and field of view (of the order of 0.001\%).
We derived the rate of coincidental binaries from our catalogues of stars, estimating a high density of 5\,000 stars in the \euclid field of view assuming a uniform distribution of the stellar positions.
As a simplification, images of binary and single stars were produced using a flat spectral energy distribution (SED). 
In real data, the stellar SEDs will be provided by the \emph{Gaia} catalogue \citep{Bruijne2012,GaiaCollaboration2016}, ground-based multi-band data, or methods such as the VIS single-image spectral classifier as proposed in \citet{Kuntzer2016b}. 
The measurements provide a set of fiducial shape parameters and sample the distortions due to binary stars. 
Mock catalogues can be drawn from these fiducial values and distortions. These catalogues contain the shape parameters ($e_1, e_2$ and the size) for each object in the observed field.

To reflect the \euclid observing strategy, four realisations of mock catalogues were prepared of the same field.
The noisy shape parameters were assumed to be known, as we described in Sect.~\ref{sec:algos}, to 1\% for the ellipticities and 5\% for the size.
The shape parameters were computed by interpolating their fiducial values from a set of the 600 \euclid PSFs.
The distortions due to a given binary system were also interpolated. 
The distortions were interpolated at their spatial position on the image, but also in terms of position of the binary with respect to its host and contrast.
The mock catalogues of the stars contained four noisy values for each shape parameter, corresponding to the four dithered observations.

After computing the interpolation, the PSF shape parameters for any system (single or binary in all its variety) can be derived and at any position.
Selection cuts were imposed on the contrast and the angular separation of the binary system during the preparation of the catalogues.
Regardless of the value of the criteria, the fraction of binaries was artificially maintained to 30\% to avoid a very low fraction of positive samples in the data, which would bias the metrics.
 
In the following numerical experiments, we only work on the catalogues.
%
\section{Numerical experiments} 
\label{sec:results}

The binary classifier algorithms were applied to two different experiments: (i) the PSF field is known to the \euclid requirements, a best-case experiment, and (ii) the PSF field is not known beforehand, the worst-case experiment. For each experiment, a training and test dataset were prepared according to Sect.~\ref{sec:mock_obs}. 
The training set was necessary in the case of the ACF method to determine the separating threshold. For RF it was used to build decision trees.

In the worst-case experiment, the training set was used to optimise the weights and biases of the ANNs and the threshold was determined on an additional validation set. 
The results reported in the following were measured on the test set that was similar to the training set, but was not shown to the algorithm during the training. 
The selection criteria (contrast and angular separation of the binaries) in both experiments evolved on a grid. The angular separations ranged between 1 and 15 mas. 
The angular separation is a lower bound criterion, while the contrast is an upper bound criterion.

%
\subsection{Known PSF field experiment}

\begin{figure}
 \begin{center}
  \includegraphics[width=1.\linewidth]{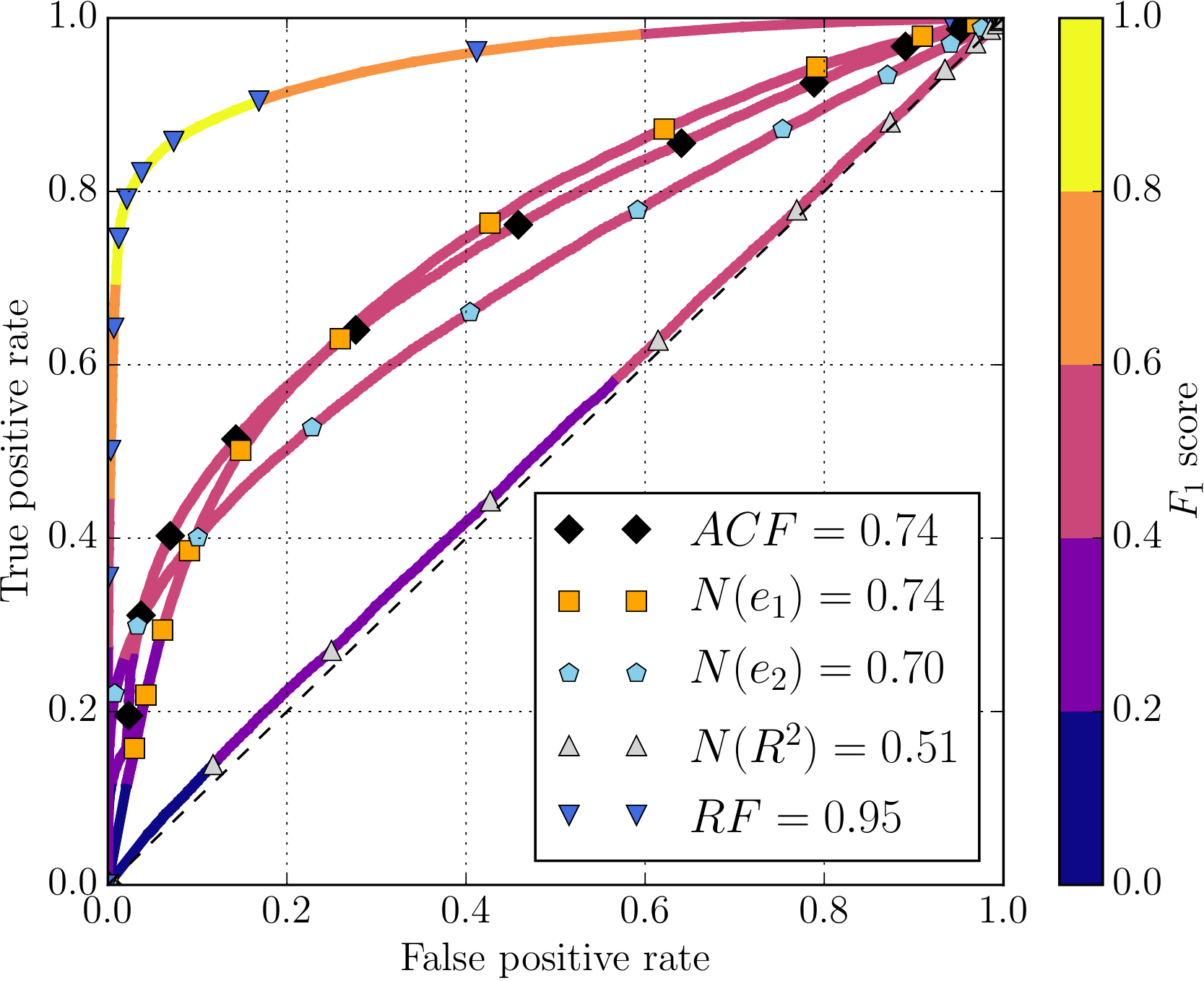}
 \end{center}
\caption{\label{fig:knPSF_roc} ROC curve for a $5$ mas minimum separation and a contrast of up to one magnitude. The curves for the single channel ($e_1$, $e_2$, and $R^2$, denoted $N(e_1)$, $N(e_2)$, and $N(R^2)$) are shown along with the combined three-channel classifier (denoted ACF) and the RF method. All curves are colour-coded with the $F_1$ score computed at each threshold. The dashed line shows the performance of a random-guess algorithm. The numerical value in the legend is the AUC of the method.}
\end{figure}
\begin{figure}
 \begin{center}
  \includegraphics[width=0.99\linewidth]{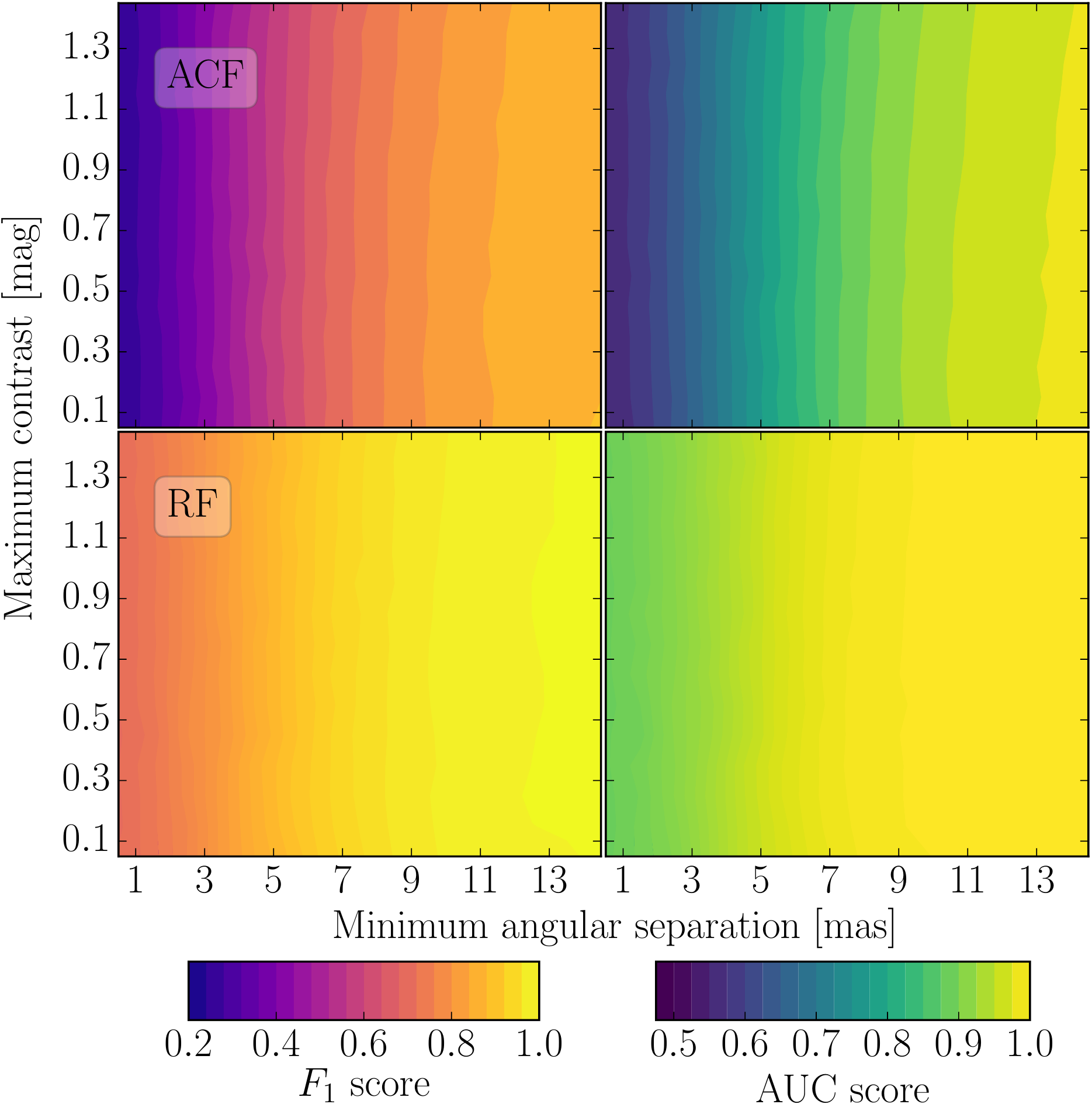}
 \end{center}
\caption{\label{fig:knPSF_perf} Performance metrics $F_1$ (left) and AUC (right) for different combinations of selection criteria in the known PSF experiment, using the ACF method (top panel) and RF (lower panel) to detect the binaries. The parameters of the PSF shapes are known to 1\% for the ellipticity and 5\% for the size.}
\end{figure}

\begin{figure}
 \begin{center}
  \includegraphics[width=1.\linewidth]{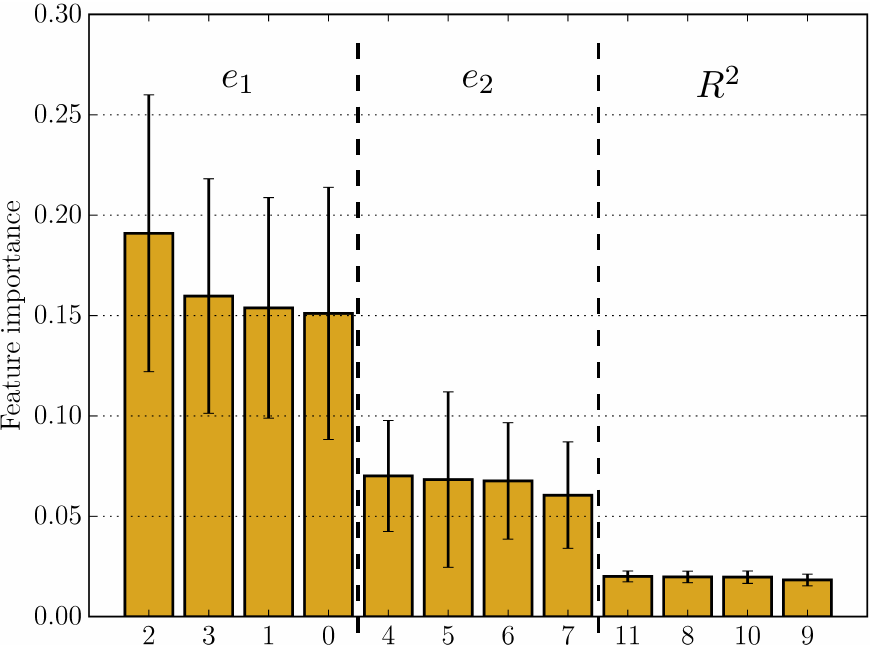}
 \end{center}
\caption{\label{fig:knPSF_featimp} Feature importance for an RF classifier with the selection criteria set at an $8$ mas minimum separation and a contrast of up to half a magnitude. Features 0 to 3 encode $e_1$ relative errors for each exposure, features 4 to 7 are the same for the $e_2$ component, and features 8 to 11 are the same for the size. The error bar shows the 1$\sigma$ deviations of the importance of the feature on the learned dataset.}
\end{figure}

In this experiment, we assumed that the PSF shape parameters are known to the \euclid requirements at any spatial position, namely $\sigma(e_i) \leq 2\times 10^{-4}$ for the ellipticity components and $\sigma(R^2)/\langle R^2\rangle \leq 1\times10^{-3}$ for the size \citep{EuclidSciReq,Paulin-Henriksson2008,Cropper2013,Massey2013}. The number of objects in the datasets depends slightly on selection criteria, but amounts on average to a few thousand objects.

For every pair of criteria, the threshold selected on the training set produced the targeted 10\% FPR in the testing phase. An ROC curve was calculated for each of the individual channels (i.e. classifying on each of the shape parameters individually) and the three-channel classifier (i.e. taking into account all shape parameters to classify). 
The ROC curves are displayed in Fig.~\ref{fig:knPSF_roc}. 
The resulting AUC, an indicator of the performance of each of the channel, was used as weight for the three-channel ACF. 
The weights were left fixed for all criteria pairs.

The most useful information to predict the nature of the object is inferred from the $e_1$ relative error features (Eq.~(\ref{eq:relerr})). The second most useful information is the second ellipticity component. While its overall performance is poorer than the $e_1$ channel, the $e_2$ channel has a better TPR at low FPR. This better performance at low FPR implies that a combination of the two ellipticity channels leads to a better overall classification. The ACF size channel classifier is a very poor predictor of binary systems. 
This remains true for most choices of selection criteria, short of very large angular separations. 

As expected, better performances in terms of both AUC and $F_1$
were obtained when the angular separation was increased. This is illustrated in Fig.~\ref{fig:knPSF_perf}. At small angular separations, the ACF classifier is barely better than a random classifier, with $F_1\sim 0.2-0.3$. At large angular separations however, the performance of the classifier is greatly improved. In general, the performance of the classifier mostly
depends on the angular separation and very little on the maximum luminosity contrast.
The separating threshold chosen was chosen on the training dataset such that the FPR was 10\%. The value of this separating threshold was the same for any choice of selection criteria.
The separating threshold was thus not influenced by the selection criteria, which simplified the application of the ACF technique. 

We then tested the RF method using the same training and test datasets as for the ACF. The RF method consistently obtained better scores than the ACF, as is shown in Fig.~\ref{fig:knPSF_perf}. The ROC curves for the RF method (Fig.~\ref{fig:knPSF_roc}) also indicate a better AUC and a higher $F_1$ score for the same FPR. 
The separating threshold is determined during the RF training to reach the goal FPR of 10\%. The relative importances of the features in the RF algorithm show, according to the ACF method, that the $e_1$ relative errors (see Fig.~\ref{fig:knPSF_featimp}) encompass the most decisive information.
The $e_2$ features come in second, while the size information is third and negligible in the decision-making process.
The RF method is only weakly dependent on the contrast, like the ACF. When the PSF field is known (to the the \euclid requirement), the RF method in particular can be used to find binary systems even at small angular separations.

%
\subsection{Unknown PSF parameter experiment} \label{sec:uknPSF}

We now turn to an experiment where we assumed no prior know\-ledge of the PSF field. To follow the procedure laid out in Sect.~\ref{sec:algos}, the PSF shape parameters must first be determined at the position of the objects of interest and the field of PSF shape parameters estimated. 
To this end, we adopted a "leave-one-out" scheme, described in algorithm~\ref{algo:leaveoneout}. The PSF shape parameters at the position of the objects are interpolated from the nearest ten and assumed single neighbours and taken as fiducial parameters. 
The shape parameters of the objects are then compared to the fiducial parameters. The resulting relative errors are then used as the vector of features for the classifiers.
The object is then classified into as a single or binary star.
At the first iteration of algorithm~\ref{algo:leaveoneout}, we assumed that all stars in the field are single. 
The above procedure was repeated to iteratively construct a good estimate of the PSF field, and to determine where the binaries are in the field of view.

\begin{algorithm}
 \begin{algorithmic}[1]
 \Procedure{UknPSF}{star\_list, $p$}
 \State  single\_stars $\gets$ star\_list 
 \For {iteration $<$ max\_iterations}
 \State  iter\_single\_list $\gets$ []
 \For {{\bf each} star {\bf in} star\_list}
        \State usable\_stars $\gets$ single\_stars $-$ star
        \State K $\gets$ Get10ClosestStars(usable\_stars, star)
        \State\# K is the list of the 10 nearest stars
        \State $w \gets$ InverseDistanceSquare(K, star)
        \State $P^{(\text{K})}\gets$MedianOverExposures($p^{(\text{K})}$)
        \State\# $P^{(\text{K})}$ is an array of K stars $\times$ 3 shape parameters, averaged over exposures
        \State $p^{(\text{star})}_{0} \gets $ WeightedAverage($P^{(\text{K})}$, $w$)
        \State\# $p^{(\text{star})}_{0} $ is a vector of the 3 shape parameters      \State $\delta^{(\text{star})} \gets (p^{(\text{star})}-p^{(\text{star})}_{0})/ p^{(\text{star})}_{0}$
        \State\# $\delta^{(\text{star})}$ is an array of 4 exposures $\times$ 3 parameters
        \If {Classification($\delta^{(\text{star})}) ==$ Single }
        \State append(star, iter\_single\_list)
        \EndIf 
 \EndFor
 \State single\_stars $\gets$  iter\_single\_list
 \EndFor
\State binary\_list $\gets$ star\_list -  single\_stars
 \State \textbf{return} binary\_list
 \EndProcedure
 \end{algorithmic}
 \caption{``Leave-one-out'' reconstruction scheme used in the case of an unknown PSF field for the ACF and RF methods in the training phase. $p$ denotes the PSF parameters for all stars and the four exposures, and $\delta$ is the relative errors.}
 \label{algo:leaveoneout}
\end{algorithm}

\begin{figure}
 \begin{center}
  \includegraphics[width=0.99\linewidth]{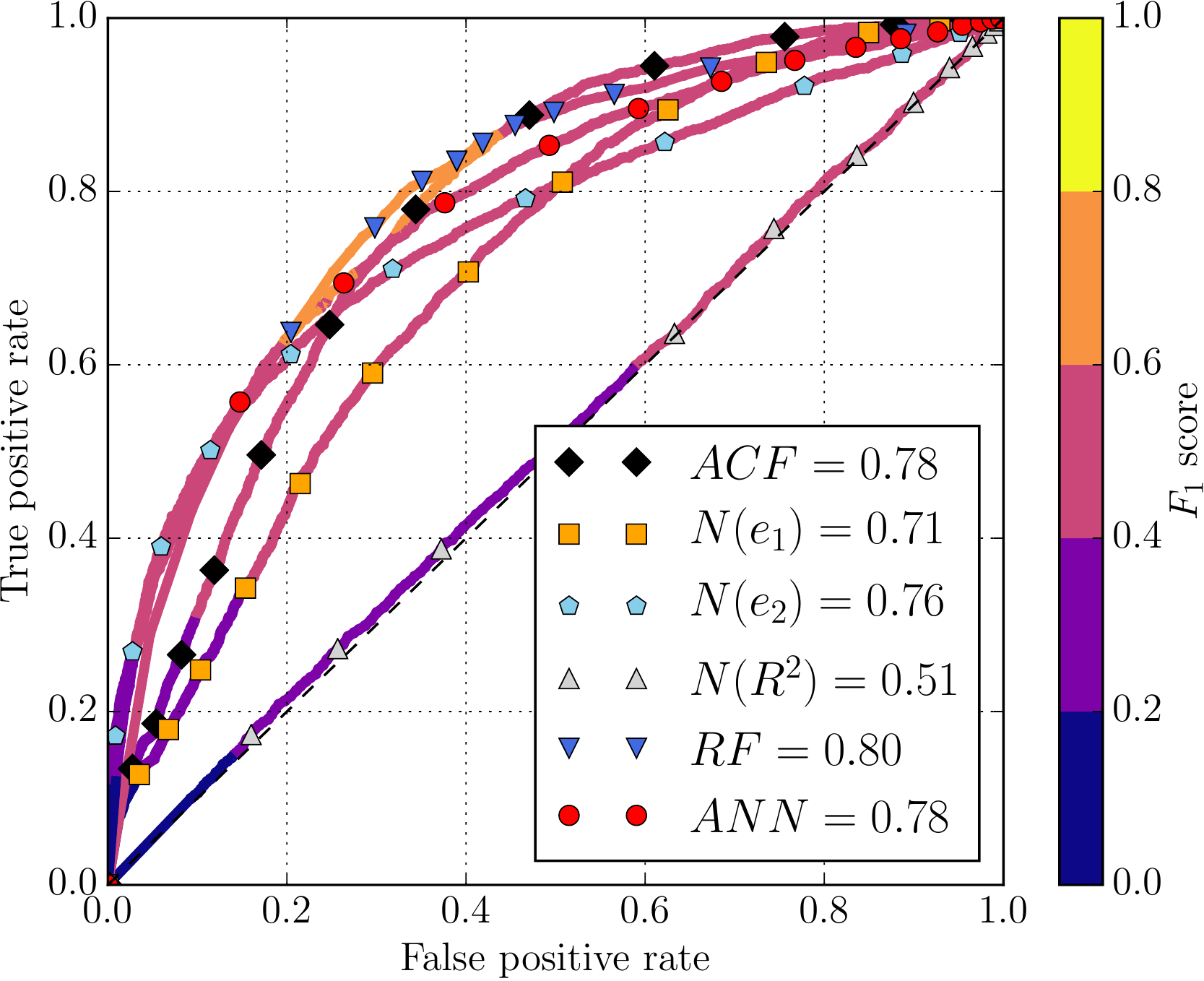} \end{center}
\caption{\label{fig:uknPSF_roc} ROC curves for the unknown PSF experiment. The angular selection criterion is 10 mas and the
contrast is of up to one magnitude. 
The curves for the single channel ($e_1$, $e_2$, and $R^2$, denoted $N(e_1)$, $N(e_2)$, and $N(R^2)$) are shown along with the combined three-channel classifier (denoted ACF), the RF, and the ANN methods. All curves are colour-coded with the $F_1$ score computed at each threshold. The dashed line shows the performance of a random-guess algorithm. The numerical value in the legend is the AUC of the method.}
\end{figure}

The VIS instrument will be comprised of 6 x 6 CCDs, each with 4k $\times$ 4k pixels, with a field of view of $0.787\times 0.709$ deg$^2$ \citep{Racca2016}. About 2\,000-3\,000 objects will be used to reconstruct the PSF field in any pointing \citep{Cropper2013,Euclid}. 
In the present work, we only have PSF estimates for the four central detectors.
The number of stars seen in this reduced field is 280, and we still use a binary fraction of 30\%. 

We determined the threshold for the ACF and inferred the decision rules for the RF on the training set in a similar way as in the known PSF field experiment. 
To test the performance of the methods, we created 20 different fields, thus different spatial positions and binary populations, each containing 280 stars. The reported metrics were averaged over the 20 fields. The relative error estimator during the testing phase was modified from Eq.~\ref{eq:relerr} to
\begin{equation}\label{eq:relerr2}
\delta^{(i)} = \frac{p^{(i)}-p^{(i)}_0}{p^{(i)}_0 + \epsilon},\end{equation}
where $\epsilon$ is a calibration coefficient accounting for the PSF field reconstruction errors.
For RF, $\epsilon_\text{RF}$ varies from 0.03 to 0.07 as a function of the selection criteria, while $\epsilon_\text{ACF}$ is in the range $[0.01,0.035]$.
If this modification is not enforced, the FPR increases several
times. 

\begin{figure}
 \begin{center}
  \includegraphics[width=0.99\linewidth]{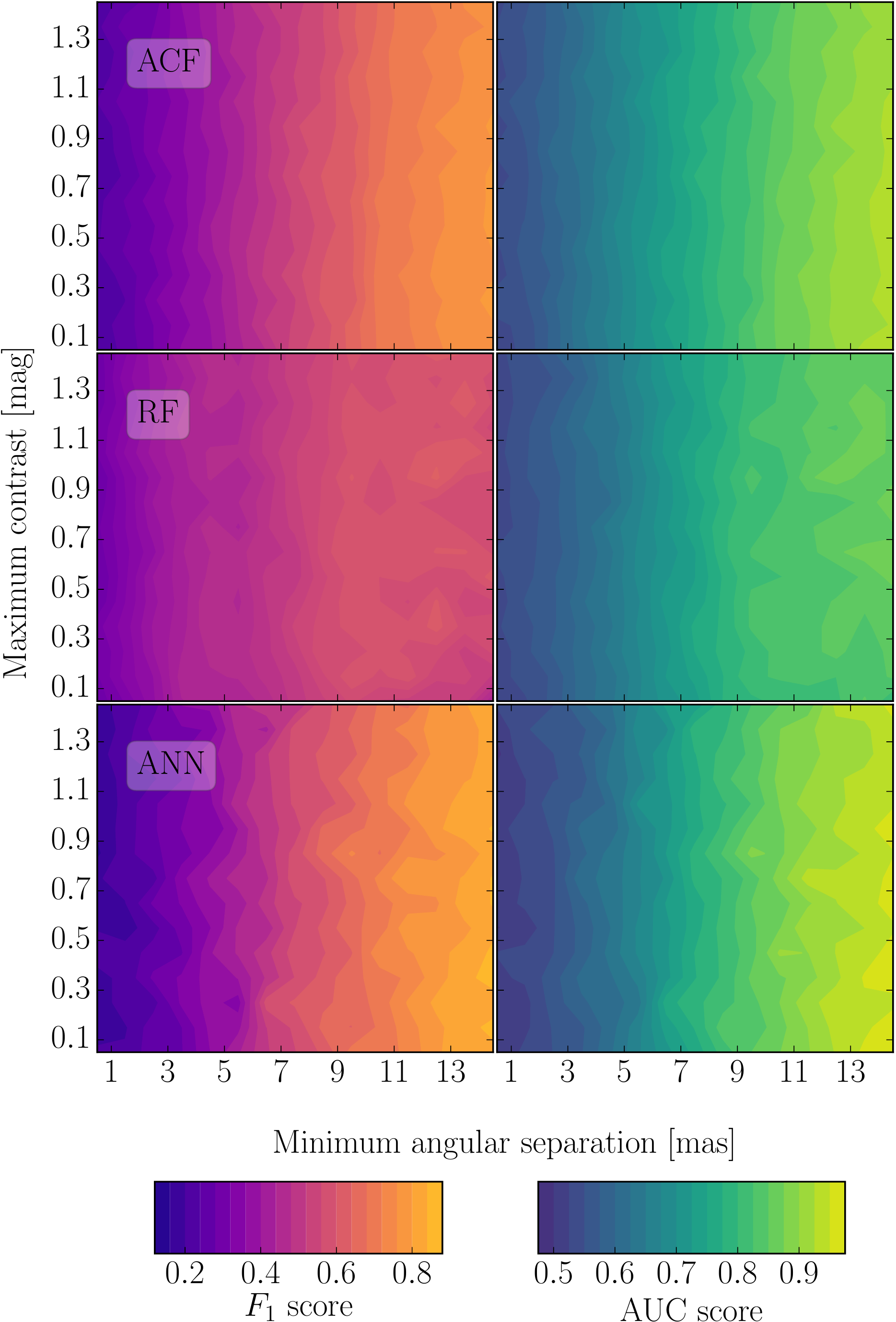}
 \end{center}
\caption{\label{fig:uknPSF_perf} Performance metrics $F_1$ (left) and AUC (right) for different combinations of selection criteria in the unknown PSF experiment, using the ACF method (top panel), RF (mid panel), and ANN (bottom panel).}
\end{figure}

The ACF method performs well in the case of an unknown PSF field. 
Even if its overall performance is reduced, it remains close to the known PSF field experiment.
Figure~\ref{fig:uknPSF_roc} shows the ROC curve in the unknown PSF case. The ACF principle is simpler in the sense that it does not need to train on the features themselves.
It only has to find a threshold between single and binary stars based on the degree of correlation of the features. 
The RF method relies on the individual relative error estimates to infer the binary nature. 
The interpolation of PSF parameters and determination of a binary star loop is not encoded in the training of the RF, which reduces the performance.
For both methods, the FPR is significantly reduced between the first iteration (in which no binaries are assumed in the field) and the second iteration to settle to about 10\% as required. After this, the FPR stays roughly constant, while the $F_1$ score tends to increase slightly.
The ACF seems to outperform the RF technique, both in terms of AUC (reaching a maximum of about 0.9 --  Fig.~\ref{fig:uknPSF_perf}) and $F_1$ score, especially at high angular separations. 
The performance of the classification again mostly depends on the minimum angular separation and not on maximum contrast. 
In Figures~\ref{fig:knPSF_perf} and \ref{fig:uknPSF_perf}, the selection is made based on maximum contrast. 
If the selection were made on minimum contrast, there would be a stronger effect.
However, we are interested in removing the most damaging objects from the PSF samples, and systems with a high contrast will not be the most damaging during the PSF reconstruction (see Fig.~\ref{fig:dang_star}).
The inclusion of the hyperparameter $\epsilon$ does succeed in producing good detection, but it is determined manually, which can hamper the image reduction pipeline. 
For this reason, we present in the next paragraph the ANN approach, which does not require the inclusion of the hyperparameter $\epsilon$.

%
\subsection{Applying ANNs to unknown PSF field experiments} \label{sec:res_ann}

The reconstruction scheme, based on the calculation of the relative errors and correlations between exposures, is subject to poor performance if there is a significant dither between exposures. The PSF can indeed spatially vary to such a degree that there is only little correlation in the PSF parameters. A more flexible approach to capture this effect is to teach the method about the deviations from the fiducial value and how it evolves with the position of the object on the chip. This latter scheme by construction handles any dithering and has the additional advantage of not having to tune an additional parameter such as $\epsilon$ in the relative error as in Eq.~\ref{eq:relerr2} to calibrate out the errors that are due to the PSF field reconstruction. 

The method was trained on parameters observed on 25 unknown fields, each containing 280 stars, with the fiducial parameters computed by the ``leave-one-out'' reconstruction scheme. We give as features the deviations from the estimated parameters and the estimated parameters. The procedure of identifying the binary star is similar to the ACF or RF procedures: first, a run on all objects, with no prior on which are binaries. Then a second pass is made, with an interpolation ignoring the objects that were blacklisted as binaries in the first run. This algorithm is almost identical to algorithm~\ref{algo:leaveoneout}, with the difference that the features used to classify are the interpolated fiducial parameters and the measured parameters. The second pass reduces the FPR by a few percentage points and slightly increases the true-positive rate. Two passes are sufficient to reach the best performance, while ACF and RF typically require three to five passes. In the current implementation, there are 15 input features (three deviations times four exposures plus the three interpolated fiducial values of the parameters) with a hidden three 15-neuron layers and 1 output neuron.
The architecture used in this work is thus: 15 inputs, 15 hidden neurons, 15 hidden neurons, 15 hidden neurons, and 1 output.
We validated the training on a set of five fields to select the separating threshold value and tested 15 fields of 280 stars.
Overall, the ANN method performs better than either ACF and RF, especially at high minimum angular separation, reaching an AUC and $F_1$ score of almost one, as shown in Fig.~\ref{fig:uknPSF_roc}. The $F_1$ score, as shown in Fig.~\ref{fig:uknPSF_perf}, is higher for most of the selection criteria. The main drawback of this method is the complexity of the training scheme.

%
\subsection{Performance dependence on the assumptions} \label{sec:assumptions}

In the following, we discuss different effects or assumptions that can significantly affect the performances of the algorithms.

\begin{itemize}
\item The measurements of the shape parameters are assumed to be accurate. 
The features should reflect a systematic bias between the exposures. If there is a systematic additive bias in the shape measurement, the algorithms could be misled into classifying single stars as binaries. 
For the ACF classifier, the performance depends on the value of the additive bias. A slow decrease in metrics value is observed until an additive bias levels of the order of $10^{-3}$. 
Past this threshold, the ACF classifier is essentially random. 
For the machine-learning classifiers, if the additive bias is included in the training, no significant degradation of the performance is noted. If the bias is not included, the same threshold in bias level is observed with the same consequences.
\item The errors on the shape measurements are assumed to be Gaussian. We tested the shape measurement algorithm based on adaptive moments on noisy PSF images at the \euclid resolution. 
Although the errors are larger than the \euclid requirements, their distribution is Gaussian. 
\item If the precision on the measurement of the size is improved from 5\% to $0.5-1\%$, the size channel carries much more weight.
If the precision of measurement on the ellipticities is improved by a factor of two from the baseline of 1\%, the performance increases by a few percent. A degradation of the same magnitude is observed if the precision is worsened by a factor of two.
 \item The number of stars per field in the \euclid survey will vary depending on Galactic latitude. The estimates range from 1,800 objects for high Galactic latitude to twice this number at low Galactic latitudes \citep{Cropper2013, Euclid}.
For the ACF, both performance metrics are stable for a number of stars per field higher than 1\,000. 
For the RF, the AUC metrics increase by 5-10\% between fields where 1\,000 stars are present and 3\,000 stars per field.
The AUC for the ANN similarly improves in the same conditions.
\item We finally discuss the completeness of the detection when the algorithms are trained with a full binary population, that is, with no selection of the binaries for training.
As expected from Fig.~\ref{fig:uknPSF_perf}, the AUC value is 0.5 and the overall $F_1$ score is poor. The completeness of the detection of binaries at large angular separation is also significantly reduced compared to methods trained on binaries selected for their large angular separation. The value of the completeness is roughly divided by two between a classifier trained to find unresolved binaries in an unknown PSF field at 10 mas minimum and a classifier trained to detect all binaries. 
\end{itemize}

%
\section{Conclusion} \label{sec:summary}

Unresolved binary stars can create significant biases on the PSF determination in space-based weak-lensing surveys like \euclid. As binary stars are ubiquitous in the sky, their observation cannot be avoided. The catalogues of single and binary stars provided by \emph{Gaia} will be useful to flag a number of undesired objects, but the binary stars identified by \emph{Gaia} are not expected to match the depth of \euclid or match its footprint on the sky \citep{Eyer2015}.

We here presented an approach to detect unresolved binaries using catalogues of shape parameters of PSFs observed multiple times.
Repeated measurements of the same objects are provided by the dithering plan and the deep field observations.

We used relative errors of the complex ellipticity and size to their fiducial value as input features for our classification algorithms. We proposed three methods. The first (ACF) is based on the auto-correlation of the relative errors of the parameters. The two others are supervised machine-learning algorithms: random forest (RF), and artificial neural networks (ANN). The methods were tested using two numerical experiments: a best case, in which the PSF parameters are known to the \euclid requirements, and a worst case, in which the fiducial PSF parameters are unknown.
Based on an analysis of the distortions caused by binary stars in a realistic \euclid setting, we suggested a detection limit for binaries separated by at least $\sim 3$~mas with a contrast lower than 1.5 magnitude to remove binaries whose distortions on the PSF ellipticity is of the order of the measurement error.

Supervised machine-learning approaches perform well in the best-case scenario, and the detection limit of 3 mas is reached. The tests on the ACF method indicate that about 50\% of the binary stars above the detection limit are correctly flagged. The influence of the contrast is weak in the detection performance, which depends mostly on the angular separation. In the worst-case experiment, the performances are degraded because the fiducial PSF shape parameters must first be determined. The RF method is limited in the worst-case scenario. ANNs performs better than the other two methods because of the increased ability of learning the relation between the deviations from the fiducial values and the binary nature of the object. The ANN approach is able to detect about 50\% of the binaries at a angular separation of 5 mas minimum. We stress that in this worse-case scenario, no prior knowledge of the PSF was used at all.

The methods proposed here are given as a proof-of-concept. Dithering between exposures is left for future implementation. The treatment of dithering can be taught to a machine-learning approach without any major change in the method, as shown by our tests with the ANN.
Priors based on the apparent spectral class of an object can be added to overcome the difficulty of the dependence of the PSF shape on the spectra of the observed point source. 
The incorporation of PSF parameters know\-ledge, even if not at the \euclid requirements, can significantly increase the effectiveness of the proposed methods.

The code corresponding to the algorithms used in this work and all scripts to reproduce the results are publicly available from a GitHub repository accessible via \url{http://lastro.epfl.ch/software}.

\begin{acknowledgements}
The authors would like to thank J\'er\^ome Amiaux, Patrick Hudelot, Koryo Okumura, and Samuel Ronayette for providing the simulated \euclid\ PSFs.
We would like to thank Henk Hoekstra and Joana Frontera-Pons for reading a draft of this paper and Vivien Bonvin for useful suggestions.
We thank the anonymous referee for their time and valuable comments.
This research has made use of NASA's Astrophysics Data System.
We are grateful to the authors of the following {\tt Python} packages: {\tt Astropy}, a community-developed core Python package for Astronomy \citep{astropy}, {\tt Matplotlib} \citep{matplotlib}, {\tt Scipy} \citep{scipy} and {\tt Numpy} \citep{numpy}.
This work is supported by the Swiss National Science Foundation (SNSF). 
\end{acknowledgements}

\bibliographystyle{aa} 
\bibliography{bib} 

\end{document}